\documentclass[journal]{IEEEtran}
\usepackage{url}
\usepackage{amsmath,amssymb,amsthm}
\usepackage[dvips]{graphicx}
\usepackage{verbatim}
\usepackage[usenames,dvipsnames]{color}
\usepackage{cite}
\usepackage{epsfig,epstopdf}
\usepackage{float}
\usepackage{algorithm}
\usepackage{algorithmic}
\usepackage[font=footnotesize]{subfig}

\newtheorem{problem}{Problem}

\makeatletter
\newcommand*{\rom}[1]{\expandafter\@slowromancap\romannumeral #1@}
\makeatother

\begin{document}

\title{Distributed Inference in the Presence of Eavesdroppers: A Survey}
\author{
    Bhavya~Kailkhura*,~\IEEEmembership{Student Member,~IEEE},
    V.~Sriram~Siddhardh~Nadendla*,~\IEEEmembership{Student Member,~IEEE},
    Pramod~K.~Varshney,~\IEEEmembership{Fellow,~IEEE}
    \thanks{
        The authors are with Department of EECS, Syracuse University, Syracuse, NY 13244.
        (Email: bkailkhu@syr.edu; vnadendl@syr.edu; varshney@syr.edu)

*These authors contributed equally to this work.
    }
}
\maketitle

\begin{abstract}
The distributed inference framework comprises of a group of spatially distributed nodes which acquire observations about a phenomenon of interest. Due to bandwidth and energy constraints, the nodes often quantize their observations into a finite-bit local message before sending it to the fusion center (FC). Based on the local summary statistics transmitted by nodes, the FC makes a global decision about the presence of the phenomenon of interest. 
The distributed and broadcast nature of such systems makes them quite vulnerable to different types of attacks. 
This paper addresses the problem of secure communication in the presence of eavesdroppers. In particular, we focus on efficient mitigation schemes to mitigate the impact of eavesdropping. We present an overview of the distributed inference schemes under secrecy constraints and describe the currently available approaches in the context of distributed detection and estimation followed by a discussion on avenues for future research.
\end{abstract}


%============================================================================================================================
%****************************************************************************************************************************
%============================================================================================================================

\section{Introduction \label{Sec: Introduction}}
Distributed inference networks have attracted much recent attention due to a variety of applications in civilian and military domains. These include surveillance, environment monitoring, cognitive radio networks and cyber physical systems.  
Distributed inference networks employ a group of sensing entities that collaborate to sense and make inferences about a given phenomenon of interest (POI). In the traditional framework of centralized inference networks, nodes transmit raw observations to the FC. These transmissions are not attractive in practice as raw observations require a large bandwidth (or energy) for reliable reception at the FC. Therefore, distributed inference networks have been proposed where the nodes transmit compressed observations which are obtained by processing original observations into a finite and tractable alphabet set.

\begin{figure}[!t]
\centering
    \includegraphics[width=3in]{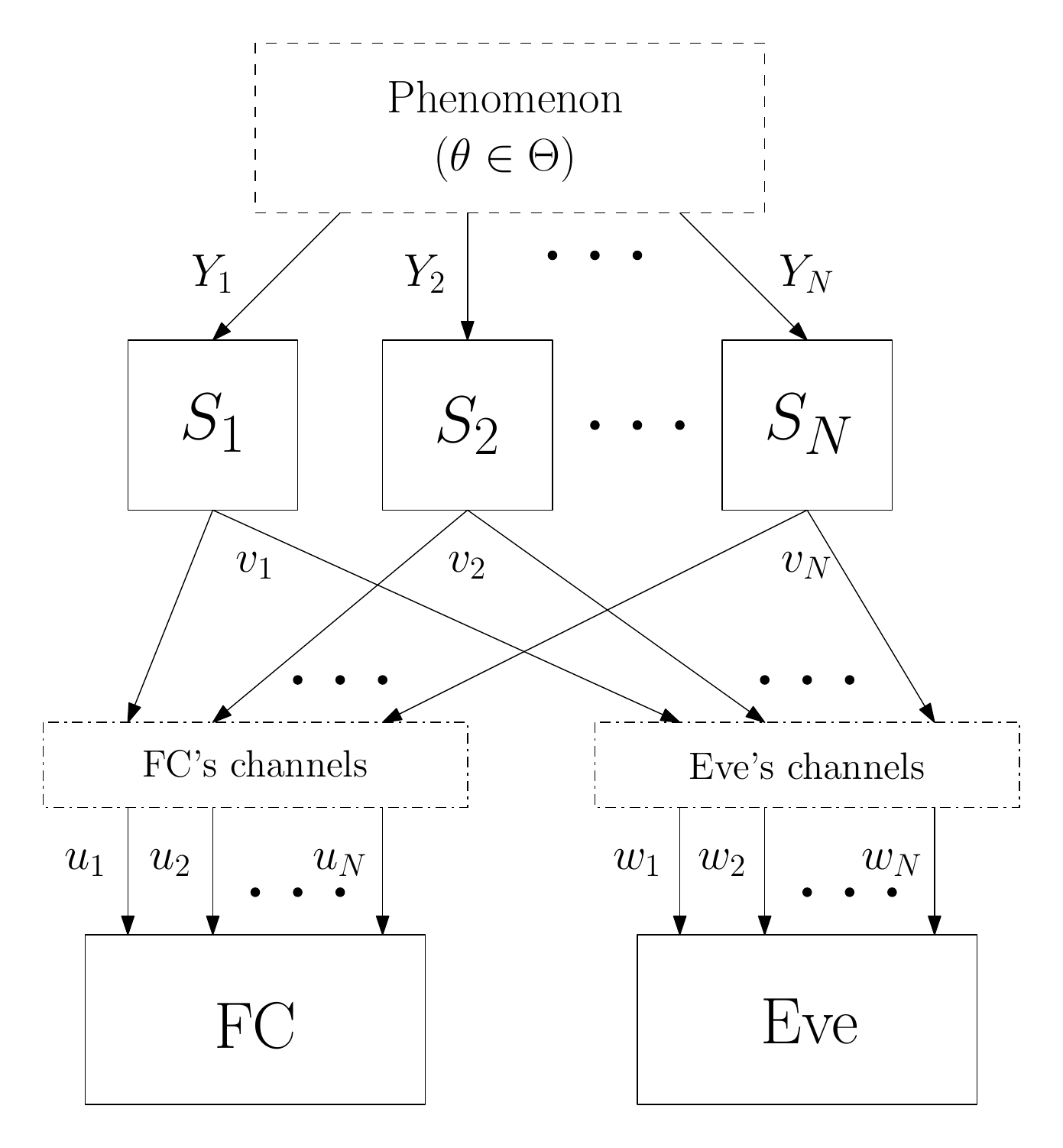}
    \caption{Distributed Inference Network in the Presence of an Eavesdropper}
    \label{Fig: model}
\end{figure}

In this paper, we denote the POI with a variable $\theta \in \Theta$, where $\Theta$ is the set of possible states that the phenomenon can take. Consider a distributed network, as shown in Figure~\ref{Fig: model}, which comprises of $N$ sensors and a central entity known as the fusion center (FC), which makes inferences about the POI. We assume that the $i^{th}$ node makes an observation $Y_i$ and compresses it into a symbol $v_i$ using a quantizer $\gamma_i$. The compressed symbol $v_i$ is then transmitted to the FC through a channel, which is represented as a function $C_F^i(\cdot)$. We denote the received symbols at the FC as $u_i = C_F^i(v_i)$, corresponding to the $i^{th}$ sensor's transmission. The FC uses the fusion rule $\Gamma_{FC}$ to integrate the symbols $\mathbf{u} = \{ u_1, \cdots, u_N \}$ into a global inference $\hat{\theta}_{FC} \in \Theta$ about the unknown phenomenon $\theta$.

Although the problem of distributed inference encompasses a broader set of problems, in this paper, we focus our attention on two fundamental problems, namely, \emph{distributed detection} and \emph{distributed estimation}. The fundamental difference in the two problems lies in the definition of the set $\Theta$. In the case of \emph{distributed detection}, $\Theta \in \{ 0, 1 \}$ and in the case of \emph{distributed estimation}, $\Theta $ is a continuous set. Practical applications of distributed detection include radar networks where the network may be interested in detecting the presence of an aircraft, or a cognitive radio (CR) network where the secondary users are interested in vacant primary user (PU) channels. On the other hand, examples of distributed estimation include location-estimation and surveillance using spatially distributed sensors.
%where privacy of the monitored entities could be breached if not addressed appropriately. \textcolor{red}{Rephrase this sentence properly.}

There are many benefits of distributed inference networks, such as bandwidth efficiency, cost effectiveness and improved reliability. However, the distributed and broadcast nature of the communication links makes the network susceptible to a breach in confidentiality. 
%\textcolor{red}{I think this is not flowing well. Needs some filler material here.} 
Thus, a breach in confidentiality of distributed inference networks is an important problem, especially when the network is a part of a larger cyber-physical system.
In a fundamental sense, there are two motives for any eavesdropper (Eve), namely \emph{selfishness} and \emph{maliciousness}, to compromise the confidentiality of a given distributed inference network. For instance, some of the nodes within a CR network may selfishly take advantage of the FC's inferences and may compete against the CR network in using the PU's channels without paying any participation costs to the network moderator. In another example, if the radar decisions are leaked to a malicious aircraft, the adversary aircraft can maliciously adapt its strategy against a given distributed radar network accordingly so as to remain invisible to the radar and in clandestine pursuit of its mission. Therefore, in the recent past, there has been a lot of interest in the research community in addressing confidentiality in distributed inference networks.

To set the notations, we represent the channel between the $i^{th}$ sensor and the Eve as a function $C_E^i(\cdot)$.
The symbol corresponding to the $i^{th}$ node received at the Eve is denoted by $w_i=C_E^i(v_i)$ (See Figure \ref{Fig: model}). In other words, the total information leakage is a function of $\mathbf{w} = \{ w_1, \cdots, w_N\}$. Similar to the FC, we assume that Eve uses a decision rule $\Gamma_E$ to integrate the symbols $\mathbf{w}$ into its own global inference $\hat{\theta}_E$. Several metrics have been proposed in the literature to quantify secrecy or the information leakage to the Eve. Some of them include  equivocation, Kullback-Leibler (KL) Divergence,  Fisher Information (FI) and probability of error. Ideally, we expect to minimize this information leakage to the maximal extent possible. For example, if KL Divergence or conditional FI is the chosen metric, then \emph{perfect secrecy} is achieved only when KL Divergence or conditional FI at the Eve becomes zero.

In this paper, we survey the state-of-the-art approaches proposed to address secrecy in the context of distributed inference networks. We first introduce a taxonomy in Section \ref{Sec: Mitigation-Approaches} where we present a survey on the state-of-the-art on secrecy in distributed inference networks. Then, in Sections \ref{Sec: Detection} and \ref{Sec: Estimation}, we specifically focus on distributed detection and estimation frameworks respectively where we present a detailed account on how secrecy is addressed in each of these frameworks. Finally, we present some important open problems while designing a secure distributed inference network in the presence of eavesdroppers in Section \ref{Sec: Open-Problems}.​

%============================================================================================================================
%****************************************************************************************************************************
%============================================================================================================================

\section{Approaches to Mitigate Threats on Confidentiality\label{Sec: Mitigation-Approaches}}
There are fundamentally four approaches to address secrecy in the context of distributed inference networks which we discuss next.
%============================================================================================================================
\subsection{Design of Sensor Quantizers and Fusion Rule \label{Sec: Mitigation - Network Design}}
In this approach, the network designer takes advantage of the difference in the channels $(C_F^i,C_E^i),$ for all $i=1,\cdots,N,$ while designing sensor quantizers and the fusion rule. We denote by $\boldsymbol\gamma = \{ \gamma_1, \cdots, \gamma_N \}$ the vector of all sensor quantizers in the distributed inference network. We assume that the quantizer $\gamma_i$ at the $i^{th}$ sensor lies within the set $\mathbb{R}_i$, for all $i = 1, \cdots, N$. Similarly, we denote the set of decision rules at the FC and Eve as $\mathbb{R}_{FC}$ and $\mathbb{R}_E$, respectively.

Without any loss of generality, we denote the performance metric at the FC and Eve as $\Omega_{FC}$ and $\Omega_E$, respectively. Consider a scenario where the network has a tolerable upper bound on the amount of information leaked to the Eve. Mathematically, this can be quantified in terms of a constraint $\alpha$ on the Eve's performance metric $\Omega_E$. Then, one way of finding the distributed inference system design in terms of sensor quantizers and the fusion rule at the FC is stated as follows.

\begin{problem}
    Find $({\boldsymbol\gamma, \Gamma_{FC}})$ such that $\Omega_{FC}$ is maximized while satisfying the constraints: 
    
    $1)$ $\max_{\Gamma_E \in \mathbb{R}_E} \Omega_E$ lies below a tolerable value $\alpha$, 
    
    $2)$ quantizers satisfy $\displaystyle \gamma_i \in \mathbb{R}_i,  \forall  i = 1, \cdots, N$, 
    
    $3)$ fusion rule at the FC satisfy $\displaystyle \Gamma_{FC} \in \mathbb{R}_{FC}$.
    \label{Problem: Network Design}
\end{problem}

Note that error exponents are asymptotic performance metrics at the FC and Eve that represent exponential decay rates of the error probability of their respective ``optimal'' detectors. Therefore, if the performance metric chosen is an error-exponent such as KL Divergence (for Neyman-Pearson detection setup) or Chernoff Information (for Bayesian detection setup), Problem \ref{Problem: Network Design} becomes independent of the fusion rules $\Gamma_{FC}$ and $\Gamma_E$ at both the FC and Eve respectively, and reduces to the design of the sensor quantizers alone.

%============================================================================================================================
\subsection{Stochastic Encryption \label{Sec: Mitigation - Stochastic Cipher}}
As an alternative to the first approach where the network is designed within the tolerable bounds on information leakage to the Eve, one can pursue a more active approach where the sensors flip their decisions randomly in order to confuse the Eve. In this case, the FC is assumed to have a better knowledge about the sensors than Eve, since the FC either deterministically knows the flipping sensors, or has knowledge about the flipping probability, about which the Eve is completely ignorant. This introduces a significant difference in the channels $(C_F^i,C_E^i),$ for all $i=1,\cdots,N,$ thus, reducing the information leakage to the Eve.

Let the alphabet set of the compressed symbols $v_i$ at the $i^{th}$ sensor be denoted as $\mathcal{A}$, where the size of $\mathcal{A}$ is denoted by $M$. In other words, the $i^{th}$ sensor employs an M-ary quantizer to compress the observation $Y_i$ into one of the $M$ symbols. Let us denote the flipping probability matrices as $\mathcal{P} = \{ P_1, \cdots, P_N \}$, where $P_i$ denotes the flipping probability matrix at the $i^{th}$ sensor which can be interpreted as pre-shared keys between the nodes and the FC. Note that $P_i$ is a stochastic matrix for any $i = 1, \cdots, N$, since all of its row elements sum up to unity. The basic problem in this case can be stated as

\begin{problem}
    Find $\mathcal{P} = \{ P_1, \cdots, P_N \}$ such that $\Omega_{FC}$ is maximized while satisfying the constraints:

    $1)$ $\max_{\Gamma_E \in \mathbb{R}_E} \Omega_E $ lies below a tolerable value $\alpha$, 

    $2)$ $P_i$ is a row-stochastic matrix, for all $i = 1, \cdots, N$.
    \label{Problem: Cipher Design}
\end{problem}

Note that, several variants of this problem can be investigated depending on the amount of knowledge the FC has regarding the stochastic encryption process. For example, one may consider that the FC has complete knowledge about the flipping probability matrices $\mathcal{P}$, however, does not know exactly whether the sensor messages are flipped or not. In this case, the FC can improve the secrecy performance at the expense of detection performance. On the other hand, the ideal scenario is the case where the FC acquires exact instantaneous knowledge regarding which sensor messages are flipped. This can be done by spending energy in the mechanism that facilitates communication between the FC and the flipping sensors. 

%============================================================================================================================
\subsection{Artificial Noise Injection\label{Sec: Mitigation - Artificial Noise}}
Another approach, similar to the case of stochastic encryption, is the addition of artificial noise to the sensor transmissions. Note that, both stochastic encryption and the addition of artificial noise to the sensor transmissions are data-falsification schemes that are employed to confuse the Eve.

In this paper, we denote the artificial noise added to the $i^{th}$ sensor's transmissions as $\eta_i$. Then, the $i^{th}$ sensor transmits $\mathbf{x}_i$ to the FC and Eve, where $\mathbf{x}_i = v_i + \eta_i$.
Let $f_i(\eta_i)$ denotes the distribution of $\eta_i$. Also, let $\mathcal{F} = \{ f_1(\eta_1), \cdots, f_N(\eta_N) \}$ denote the set of artificial noise distributions employed by all the sensors in the network. Then the problem can be stated as follows.

\begin{problem}
    Find $\mathcal{F} = \{ f_1(\eta_1), \cdots, f_N(\eta_N) \}$ such that $\Omega_{FC}$ is maximized while satisfying the constraints:

    $1)$ $\max_{\Gamma_E \in \mathbb{R}_E} \Omega_E $ lies below a tolerable value $\alpha$, 

    $2)$ $f_i(\eta_i)$ is a probability density function of $\eta_i$, for all $i = 1, \cdots, N$.
    \label{Problem: Artificial Noise}
\end{problem}

%============================================================================================================================
\subsection{MIMO Beamforming \label{Sec: Mitigation - Beamforming}}
In order to ensure minimal performance loss at the FC as a trade-off to attaining the secrecy constraint at the Eve, another alternative approach is to use MIMO beamforming, where the sensor messages are directed towards the FC. In this case, we assume that the sensors are equipped with multiple antennas to transmit their messages to the FC. The beamforming mechanism is designed in such a way that some of the available energy is invested in the beams directed towards the FC, while the nulls towards the Eve.

In this paper, we denote the number of antennas at the $i^{th}$ sensor as $L_i$. Therefore, the $i^{th}$ sensor constructs a vector $\mathbf{x}_i$ based on the symbol $v_i$ and transmits it to the FC and Eve, respectively. Based on the channel gains at the FC and Eve, this $\mathbf{x}_i$ is designed to appear very noisy at the Eve, and simultaneously have significant information about the compressed symbol $v_i$ at the FC. For example, let $\mathbf{x}_i$ be constructed as $\mathbf{x}_i = \mathbf{b}_i v_i$,
where $\mathbf{b}_i$ is the beamforming gain vector of the $i^{th}$ sensors' signal. Assuming that both the FC and Eve have only a single antenna, the resulting received symbol at the FC and Eve are given by $u_i$ and $w_i$ respectively. Let $n_{FC_i}$ and $n_{E_i}$ denote the noise at the FC and Eve respectively. Then, $u_i  =  \displaystyle \mathbf{h}_i^T \mathbf{x}_i + n_{FC_i} = \displaystyle v_i \mathbf{h}_i^T \mathbf{b}_i + n_{FC_i}$ and $w_i = \displaystyle \mathbf{g}_i^T \mathbf{x}_i + n_{E_i} = \displaystyle v_i \mathbf{g}_i^T \mathbf{b}_i + n_{E_i}.$
Let the beamforming matrix be denoted as $B = \displaystyle [ \mathbf{b}_1 \ \cdots \ \mathbf{b}_N ]$. Since, any practical sensor is energy-constrained, we assume that the total energy available at the $i^{th}$ sensor is denoted by $E_i$. Then, the design problem can be formally stated as follows.

\begin{problem}
    Find $B = \displaystyle [ \mathbf{b}_1 \ \cdots \ \mathbf{b}_N ]$ such that $\Omega_{FC}$ is maximized while satisfying the constraints:

    $1)$ $\max_{\Gamma_E \in \mathbb{R}_E} \Omega_E $ lies below a tolerable value $\alpha$,

    $2)$ $\mathbf{b}_i$ is chosen such that the total transmit energy is within the prescribed limit $E_i$, for all $i = 1, \cdots, N$.
    \label{Problem: MIMO beamforming}
\end{problem}

Note that, all of the above approaches can be combined together to design system in a holistic manner and attain a better performance in terms of $\Omega_{FC}$, given a tolerable Eve's constraint $\alpha$. 

%============================================================================================================================
%****************************************************************************************************************************
%============================================================================================================================

\section{Secrecy in Distributed Detection\label{Sec: Detection}}
In this section, we provide a survey on the state-of-the-art on how secrecy is addressed within the framework of classical and compressive detection networks respectively. In both these frameworks, we organize the survey according to the four different approaches listed in Section \ref{Sec: Mitigation-Approaches}. 

\subsection{Classical Distributed Detection}
First, we focus our attention to the first approach where the distributed detection network (i.e., sensor quantizers and fusion rule) is optimized while satisfying the secrecy constraints at the Eve. Nadendla \emph{et al.}, made the first attempt in the year 2010 in \cite{sidhao} where they considered an unconstrained differential secrecy problem. Let us denote KL Divergences at the FC and Eve by $D_{FC}$ and $D_E$, respectively. Now, Problem \ref{Problem: Network Design} in their setup reduces to the design of sensor quantizers alone, with $\Omega_{FC}=D_{FC}-D_E$ and $\alpha=\infty$. It was assumed that the channel-state information is completely known at both the FC and the Eve. The authors showed that in the case of eavesdropper with noisier channels, the optimal local detectors are always on the boundaries of the achievable region of sensor's ROC and, therefore, are likelihood-ratio tests (LRTs). Later, the authors also considered Problem \ref{Problem: Network Design} with $\Omega_{FC} = D_{FC}$ and $\Omega_E = D_E$, in which case, the structure of an optimal local detector was conjectured to be a LRT-based test based on numerical results. 

Marano et al.~\cite{marano}, in 2009, considered the problem of designing optimal decision rules for a sensor network where the sensors perform censoring in order to save energy. It was assumed that the eavesdropper does not have access to the sensors' transmitted data but can monitor the transmission activity of the channel and exploit the busy/idle state of the channel for detecting the hypothesis. KL Divergence was used as the performance metric for both the FC and the Eve, and a censoring strategy is developed in order to maximize the divergence of FC while ensuring that the divergence of Eve was zero (perfect secrecy). Although their framework of censoring sensor networks is more general, they assumed that the Eve can only determine whether an individual sensor transmits its decision or not. In reality, Eve can extract more information than just merely determining the presence or absence of transmission, and hence can make a reasonably good decision based on its receptions.

Li \emph{et al.}, in 2014, investigated the problem of Bayesian distributed detection with two nodes in the network in the presence of an eavesdropper in \cite{bayes}, where the Eve has access to only one of the sensor's transmissions. Here, $\Omega_{FC}$ and $\Omega_E$ were assumed to be negative expected detection costs at the FC and the Eve respectively. The authors proved that LRT-based tests were optimal at the sensors if the network is designed to minimize the expected detection cost at the FC such that the minimum average cost at the Eve is no greater than a prescribed non-negative value $\alpha$.

\begin{figure*}[!t]
    \centerline
    {
        \subfloat[Sensor's ROC in the presence of Eve]{\includegraphics[width=3.3in]{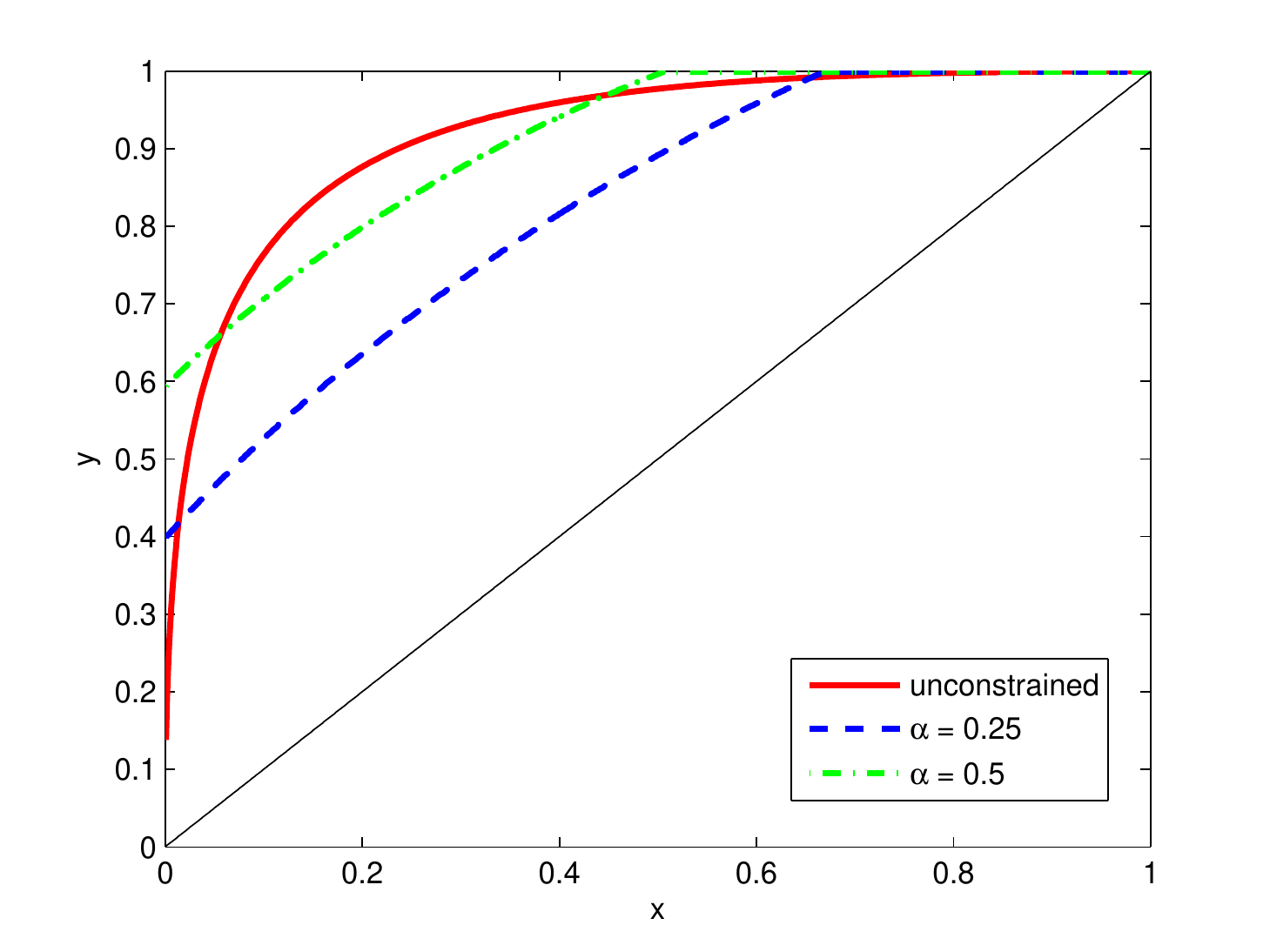}%
        \label{Fig: ROC-constrained}}
        \hfil
        \subfloat[$D_{FC}$ as a function of local false-alarm probability]{\includegraphics[width=3.3in]{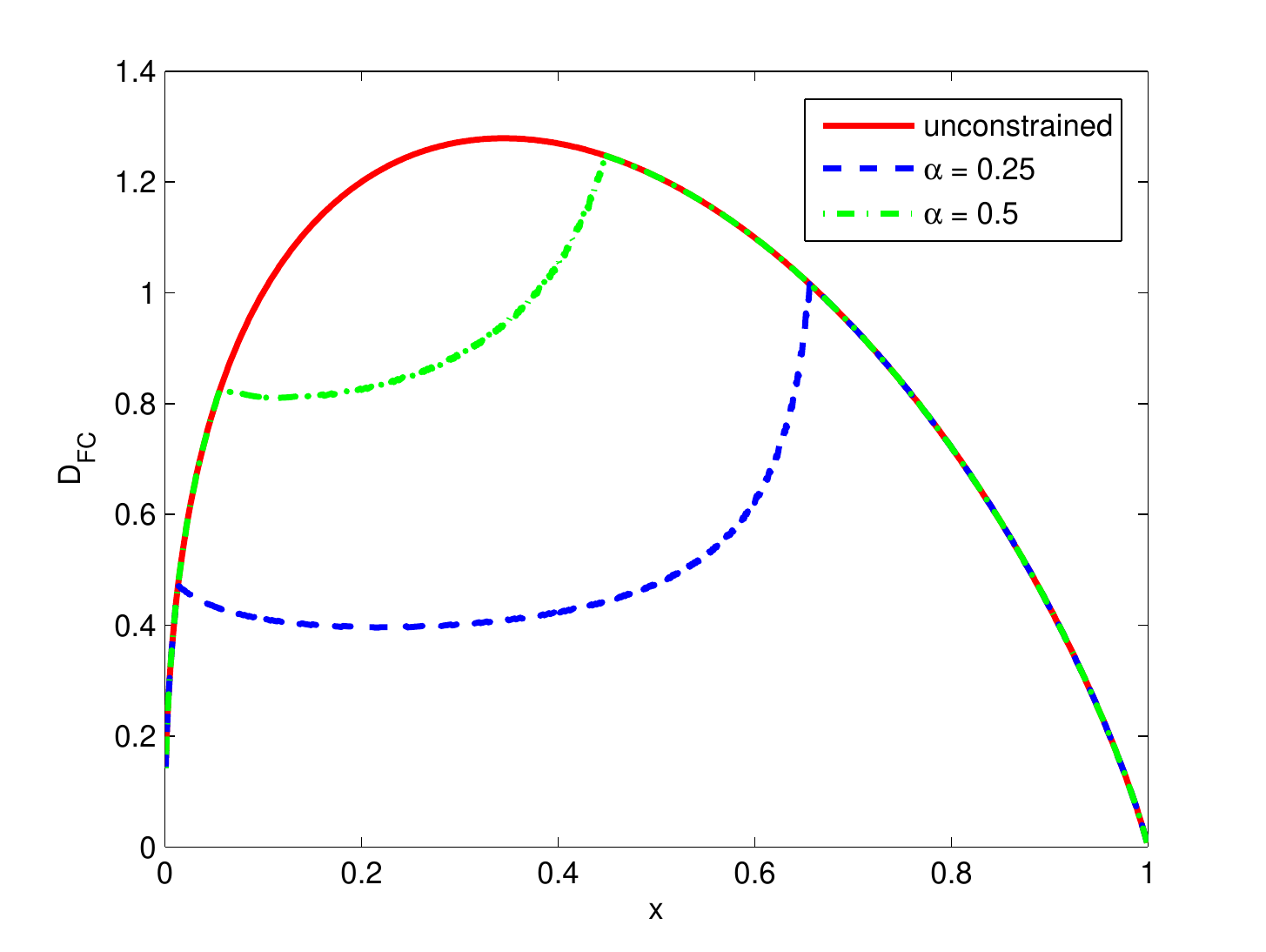}%
        \label{Fig: D_FC-constrained}}
    }
    \caption{Sensor performance in the presence of a constraint, $D_E \leq \alpha$, where $\rho_e = 0.1$~\cite{sid-2014}.}
    \label{Fig: Eve-constrained}
\end{figure*}

Li \emph{et al.}, also investigated the detection problem under the Neyman-Pearson setup for the same network as in \cite{np}.
The sensor quantizers and the fusion rule were designed to maximize the FC's probability of detection ($\Omega_{FC}$) in the presence of constraints on false-alarm probabilities at the FC and Eve, along with the probability of detection at the Eve ($\Omega_E$). Note that the false-alarm constraints at both the FC and the Eve are captured by the feasibility sets $\mathbb{R}_{FC}$ and $\mathbb{R}_E$ respectively. Here, the authors proved that the optimal local quantizer is a deterministic LRT, while the fusion rule may still be a randomization between two or more LRTs. Later, in 2014, Nadendla \emph{et al.} investigated a more general framework in \cite{sid-2014} with $N$ sensors. Here, they proved the conjecture stated in \cite{sidhao} in the context of binary symmetric channels between the sensors, FC and Eve. An algorithm was also presented to find optimal thresholds for the likelihood-ratio quantizers when the sensor observations are corrupted by additive Gaussian noise. Figure \ref{Fig: Eve-constrained} depicts the behavior of FC's performance in terms of both receiver operating characteristics (ROC) and KL Divergence at the FC as a function of tolerable limits on Eve's KL Divergence. Note that, the optimal quantizer is always on the intersection of the ROC and the Eve's constraint curve. The authors also showed that the network with non-identical sensors and channels can be designed by solving $N$ sequential problems, where the order of this sequence is dictated by the quality of the corresponding sensor's channel.

%\textcolor{blue}{Bhavya: I think we should order all the works in a chronological order. Otherwise, the story seems distorted.}

Next, we survey the literature that addresses the second mitigation approach where a stochastic cipher is employed to confuse the Eve regarding the true phenomenon. In~\cite{opt}, Soosahabi \emph{et al.} employ J-divergence as the performance metric for both the FC and Eve and design a network that guarantees perfect secrecy. This is achieved by fixing $\alpha = 0$ in Problem \ref{Problem: Artificial Noise}. Probabilistic ciphers were also studied in~\cite{nagpor} where the performance metric chosen was the error probability in the case of both FC and Eve. Note that both \cite{opt, nagpor} assume the existence of an underlying key-exchange mechanism that is secure from Eve. Alternatively, channel-aware stochastic ciphers use seeds that are obtained by exploiting randomness in the channel-gains between the node and the FC. For example, Jeon \emph{et al.}, in~\cite{stbma}, proposed a type-based multiple access (TBMA) protocol for a distributed detection network with a multiple access channel (MAC). Here, some of the nodes in the network are selected to deliberately transmit interfering signals so as to minimize degradation in the FC's detection performance, while simultaneously preventing Eve from identifying the sensors generating interference. Note that the above scheme requires full channel-state information at the sensors, and therefore, may be impractical in some scenarios. In order to alleviate this problem, efforts such as \cite{jeon2} have been made in the literature, where Jeon \emph{et al.} designed a secure transmission strategy for the local nodes in a parallel distributed detection network, where the FC first broadcasts known symbols and two thresholds to let the nodes measure their channel condition. Depending on the received symbols, the nodes are divided into three groups, non-flipping, flipping, and dormant groups. The non-flipping set of sensors quantize the sensed data and transmit them to the FC, while the flipping sensors transmit flipped decisions in order to confuse the Eve. The sensors within the dormant set sleep, in order to conserve energy and have an energy-efficient sensor network with longer lifetime.

Finally, there have been efforts to design a hybrid mitigation approach that combines the effects of both the first and the second approaches. In this regard, in \cite{Nadendla-MSThesis}, Nadendla considered the problem of Bayesian distributed detection in the presence of an eavesdropper, where the nodes use identical threshold quantizers to make their binary decisions and encrypt them before transmission using a simple probabilistic cipher. 
Cipher parameters and threshold were optimized jointly so as to ensure an acceptable probability of error at the FC while maximizing the probability of error at the Eve. 

\subsection{Collaborative Compressed Detection}
In scenarios where the POI is a high dimensional signal vector, the Collaborative Compressed Detection (CCD) framework has been proposed. 
In contrast to conventional detection framework, in CCD, the detection problem is solved directly in compressive measurement domain.
More specifically, the CCD framework comprises of a group of spatially distributed nodes which acquire observations regarding the high dimensional ($K\times 1$) signal vector to be detected. Nodes compress their observations using a $M \times K$ low dimensional $(M<<K)$ random projection operator $\phi$. Each node $i$ sends an un-quantized (or quantized) version of compressed observation vector $Y_i$ to the Fusion Center (FC) where a global decision is made.

First, we focus our attention on the first approach where  nodes do not quantize their observations and the FC receives compressed observation vectors, $\mathbf{Y}=[Y_1,\cdots,Y_N]$.
Kailkhura et al. in~\cite{asilomar} considered the problem of collaborative signal vector detection using un-quantized compressive measurements under a physical layer secrecy constraint $\Omega_E\leq \alpha$.
To counter Eve, the authors proposed to use $\beta$ fraction of cooperative nodes that assist the FC by injecting artificial noise (adding or subtracting a constant vector $D_i$ from their observation vector $Y_i$) in the system to confuse the eavesdroppers. The authors employed deflection coefficient, $d_i$, as the performance metric for both the FC and the Eve, thus, $\Omega_{FC}=d_{FC}$ and $\Omega_E=d_E$. The problem of determining optimal system parameters (i.e., compression ratio $c$ and noise injection parameters $(\beta, D_i)$) which maximize $d_{FC}$, while ensuring perfect secrecy at the eavesdropper (information of the eavesdropper is exactly zero, i.e., $\alpha=0$) was also considered. 

Kailkhura et al. in~\cite{wrkshp} extended the CCD framework to the case where compressive measurements were quantized to one-bit using LRT. The performance metric was assumed to be the probability of error $P_E$. They proposed to use $B$ out of $N$ cooperating trustworthy nodes that assist the FC by providing flipped decisions (stochastic enciphering with $P_i = \left( \begin{array}{cc} 0 & 1 \\ 1 & 0 \end{array} \right)$ for all $i = 1, \cdots, B$) to the Eve to achieve perfect secrecy. The authors considered the problem of designing optimal system parameters (fusion rule, compression ratio $c$ and fraction of data falsifying nodes $\beta=B/N$) such that $P_E$ at the FC is minimized while ensuring perfect secrecy.
\begin{figure}[t!]
  \centering
    \includegraphics[width=3.25in]{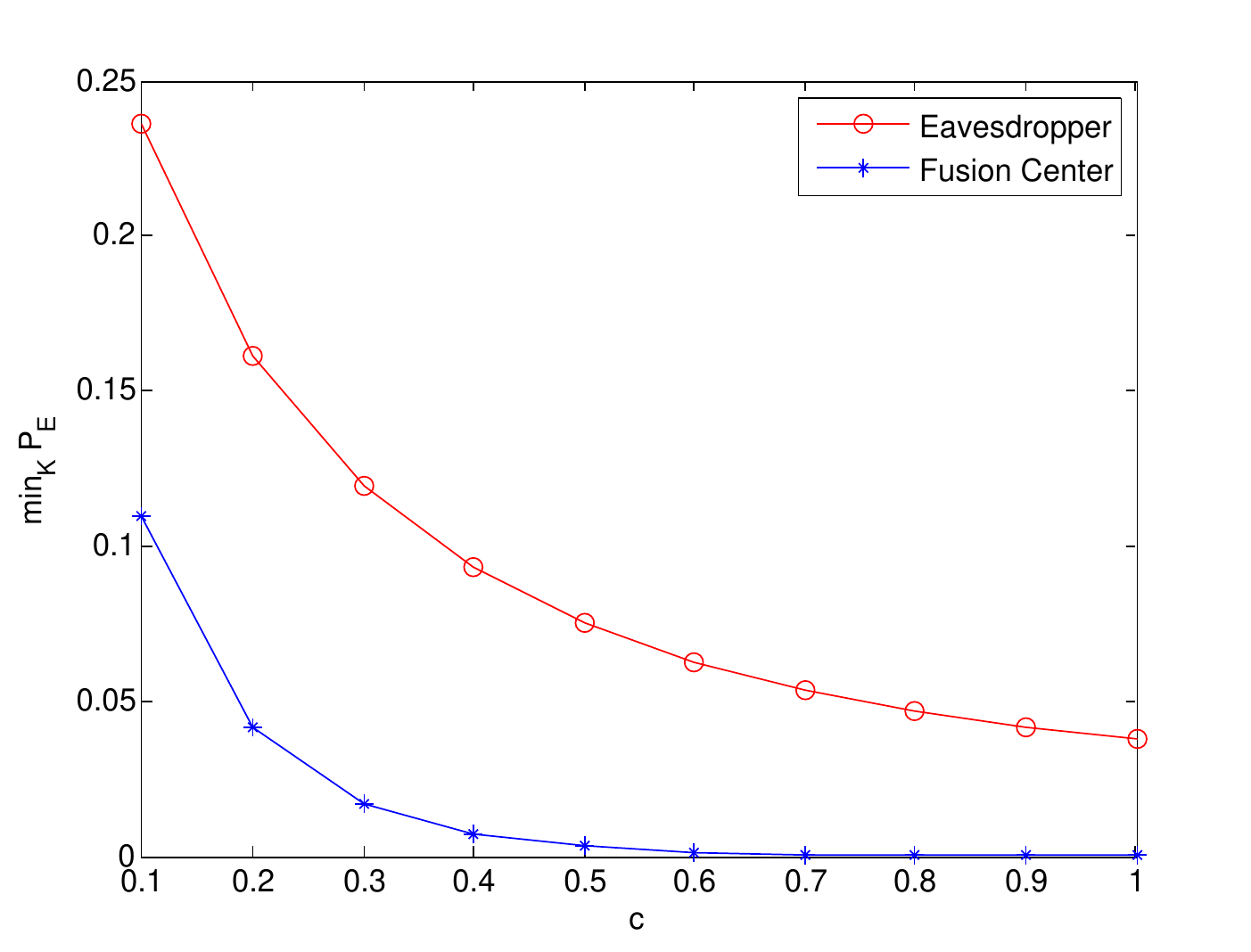}
    \caption{Minimum Probability of Error $\min_K P_E$ as a function of compression ratio $c$ where local sensor threshold $\lambda=1$, $\beta=0.2$, SNR$=10$dB and $N=10$~\cite{wrkshp}.}
    \label{tr1}
\end{figure}
In Figure~\ref{tr1} the minimum probability of error (for equal prior case), both at the FC and at the eavesdropper, is plotted as a function of compression ratio $c$. It can be seen from Figure~\ref{tr1} that the detection performance, both at the FC and at the eavesdropper, is a monotonically increasing function of the compression ratio, i. e., detection performance is better with less compression. This suggests that compression improves security performance at the expense of detection performance.

%The authors in~\cite{tandem} considered the problem of Bayesian distributed detection with an eavesdropper in tandem topology. They considered a simple network with three tandem-connected nodes %$S_1$, $S_2$, and $S_F$ (the fusion node), and an eavesdropper. Each node makes a binary decision based on the decision of the previous node (if available) and his own observation which is corrupted by the sensing channel noise. It was assumed that sensing channels are independent and their likelihood ratios contain no point masses of probability. The eavesdropper is supposed to overhear the link $S_1-S_2$ (Scenario A) or $S_2-S_F$ (Scenario B) and to make its own binary decision based on the overheard local decision. Similar results as in~\cite{bayes} for Bayesian distributed detection with an eavesdropper in parallel topology was derived, i.e., it is sufficient to consider likelihood-ratio tests for detection at all nodes including the FC.

%============================================================================================================================
%****************************************************************************************************************************
%============================================================================================================================

\section{Secrecy in Distributed Estimation \label{Sec: Estimation}}

In this section, we survey the state-of-the-art on how breaches in confidentiality are mitigated in distributed estimation networks. Although little work has been published that addresses secrecy in the context of distributed estimation when compared to the richer literature on secrecy in distributed detection, we again focus on each mitigation technique presented in Section \ref{Sec: Mitigation-Approaches}.

\begin{figure}[t!]
  \centering
    \includegraphics[width=3.5in]{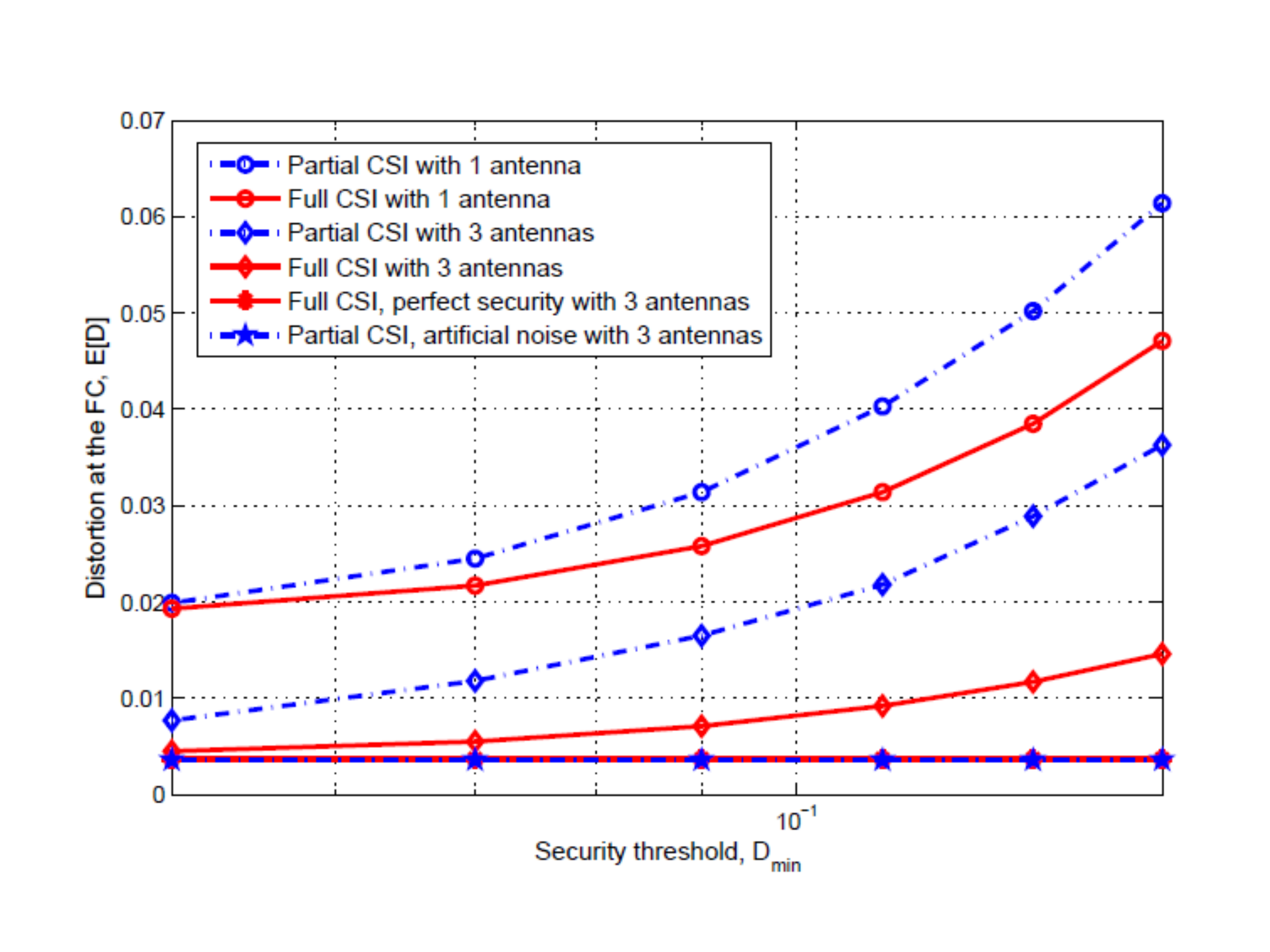}
    \caption{Performance comparison between full CSI, partial CSI and artificial noise in a multiple-antenna system~\cite{secest}.}
    \label{tr4}
\end{figure}

First, we survey the first approach in the context of distributed estimation networks where the sensor quantizers and the fusion rule are designed to guarantee the tolerable limits on Eve's performance. For example, Guo \emph{et al.}, in \cite{secest}, considered the problem of estimating a single point Gaussian source in the presence of Eve, where the sensor observations are transmitted using an amplify-and-forward technique over a slow-fading orthogonal MAC. Two different scenarios have been addressed within this framework: one, where there are multiple nodes, with each node having a single transmit antenna, and another scenario where a single node has multiple antennas. Through appropriate power allocation at the sensors, the network is designed to achieve the minimum mean squared error (MSE) regarding the POI in each of the above mentioned scenarios while guaranteeing MSE at the Eve  to be greater than a threshold $\alpha$. As shown in Figure~\ref{tr4}, the authors plot the distortion (MSE) performance at the FC with respect to the security threshold $\alpha=D_{min}$, with the transmission power budget being set to 30mW, for a one-antenna case and a three-antenna case respectively. For comparison, the system performance is depicted under four settings, namely partial CSI, full CSI, full CSI with perfect secrecy, and partial CSI with artificial noise. First, due to the channel knowledge of both the FC and the Eve, it is not surprising to see that the performance of full CSI scenario is superior to the performance of partial CSI, and the gap keeps increasing as we increase the secrecy threshold. Another important observation is the small gap between the MSE in the perfect secrecy setting and the MSE in the setting with artificial noise. A similar performance was also obtained for the multiple nodes network, where each node has only one transmit antenna.

\begin{figure}[t!]
  \centering
    \includegraphics[width=3.5in]{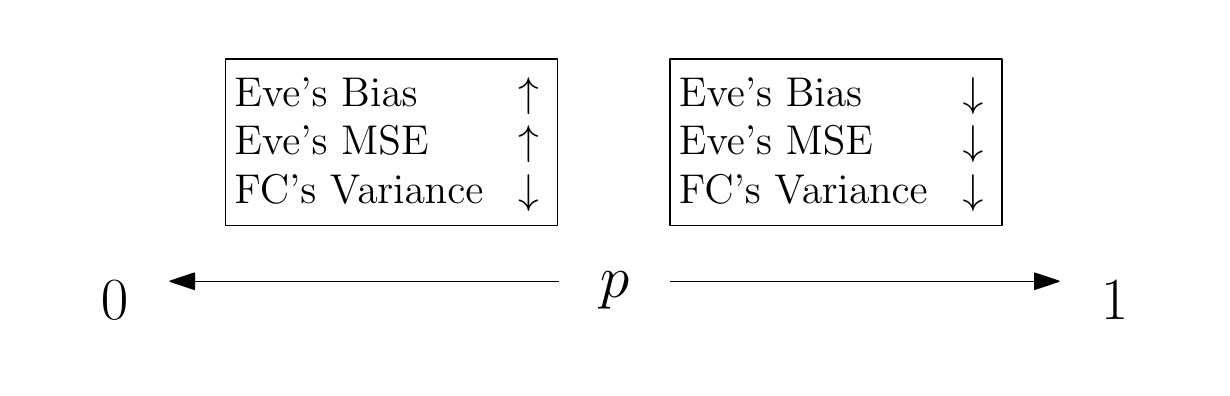}
    \caption{Effect of varying $p$ on the FC's CRLB (variance of the optimal ML estimator) and the bias and MSE of the Eve~\cite{aysal}.}
    \label{tr2}
\end{figure}

\begin{figure}[t!]
  \centering
    \includegraphics[width=3.5in]{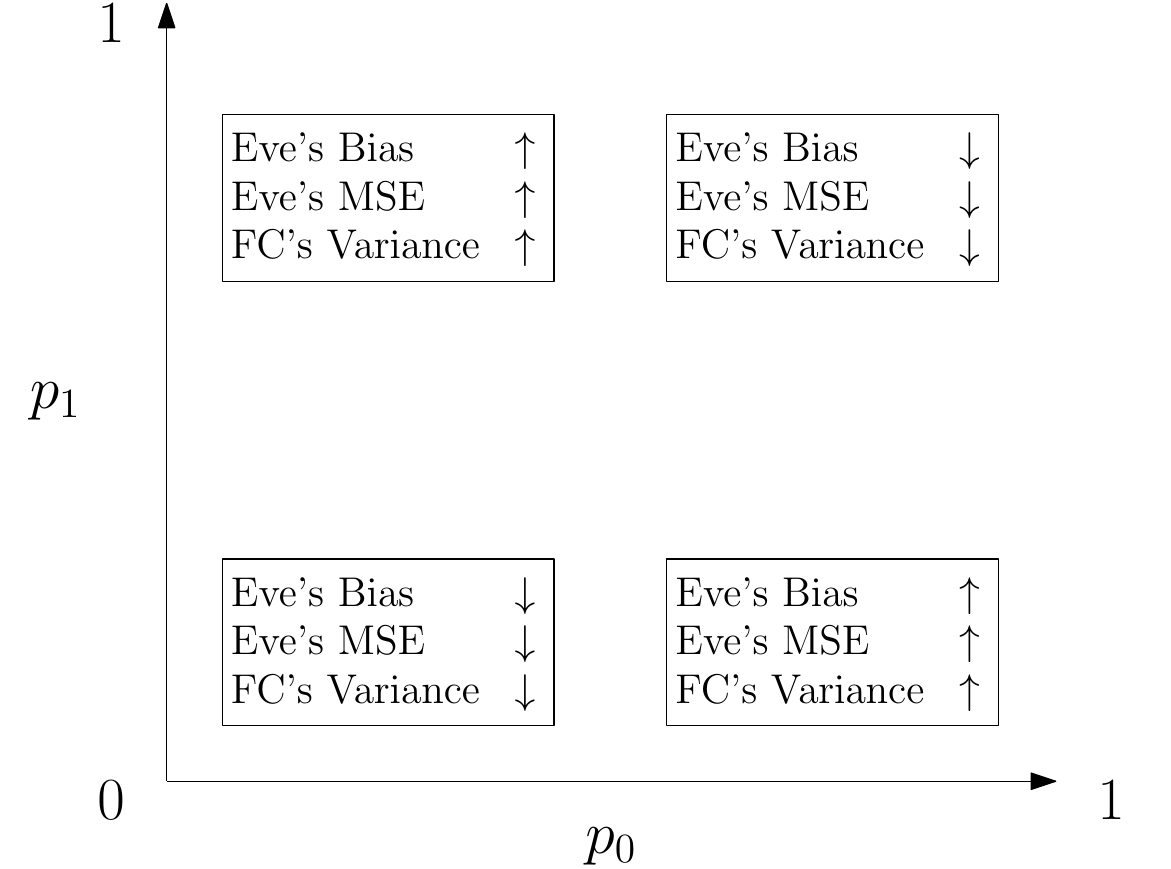}
    \caption{Effect of varying $p_0$ and $p_1$ on the FC's CRLB, Eve's bias and Eve's MSE~\cite{aysal}.}
    \label{tr3}
\end{figure}

Next, we survey how stochastic encryption is used to achieve the secrecy guarantees within the framework of distributed estimation. Aysal \emph{et al.}, in \cite{aysal} considered the problem of distributed estimation of a deterministic signal in the presence of an Eve, where each node collects a noisy observation, performs binary quantization, and transmits the 1-bit decision to the FC. The authors assume that both the FC and Eve pursue maximum-likelihood estimation in the presence of a stochastic cipher, for which bias, variance, and MSE were derived in closed form. In the context of symmetric ciphers where $P_i = \left( \begin{array}{cc} 0 & p \\ p & 0 \end{array} \right)$ for all $i = 1, \cdots, N$, the behavior of Eve's bias and MSE and FC's CRLB are characterized in Figure \ref{tr2}. Note that, as $p \rightarrow 0$: 1) the Eve's bias increases; 2) the Eve's MSE increases; and 3) the CRLB decreases. On the other hand, as $p$ tends to unity: 1) the Eve's bias decreases, 2) the Eve's MSE decreases, and 3) the CRLB decreases. In other words, choosing a smaller $p$ is better as it results in a significant amount of bias and MSE at the Eve, with a marginal increase in the estimation variance at the FC. In the case where $P_i = \left( \begin{array}{cc} 0 & p_0 \\ p_1 & 0 \end{array} \right)$ for all $i = 1, \cdots, N$ with $p_0 \neq p_1$, the effect of varying $p_0$ and $p_1$ on the FC's CRLB, Eve's bias and Eve's MSE are summarized in Figure~\ref{tr3}. In their numerical results, the authors also demonstrated that asymmetric ciphers (i.e., ciphers with asymmetric flipping probability matrices) produce greater bias and MSE than the symmetric ciphers.

%============================================================================================================================
%****************************************************************************************************************************
%============================================================================================================================

\section{Summary and Open Problems \label{Sec: Open-Problems}}
Despite the increasing attention on the problem of secure distributed inference in the presence of eavesdroppers, research in this area is still at an early stage. So far, four different approaches have been proposed to mitigate breaches in confidentiality in the context of distributed inference networks. 
But, all of these four approaches rely on an important underlying assumption that Eve's channels $C_E^i,$ for all $i=1,\cdots,N,$ are completely known at the FC and vice-versa, which may not be true in practice. In fact, there have been no works in the context of inference networks on how one can acquire the information about a  passive Eve's channel. This is a hard problem to solve because there is no feedback from the Eve to any of the nodes in the network regarding its presence or activity. An alternative to this roadblock is to assume that Eve's channel belongs to a set $\mathcal{C}$, and, investigate the best and the worst case performance at the Eve over a class $\mathcal{C}$. Information regarding this set $\mathcal{C}$ can be obtained from the scene where the network is deployed.

Also, the designers may extend the aforementioned four fundamentally different approaches into several hybrid approaches by considering two or more of these approaches together to create a more sophisticated and improved system in terms of FC's performance for a given tolerable constraint on Eve. Although there have been a few attempts in this direction, one can still envision many such hybrid mechanisms where the designer may accumulate the benefits of each of these approaches. Of course, there is always a need for any new approach which is fundamentally different from any of the four approaches listed in this paper.

\bibliographystyle{IEEEtran}
\bibliography{Journal}

\end{document}